\documentclass[12pt]{article}
\usepackage{amscd,amsmath,amssymb,amstext,amsthm,exscale,latexsym}
\newcommand {\g }   {\gamma}       
\newcommand {\dl}   {\delta}

\newcommand {\lm}   {\lambda}      
          
\newcommand {\s }   {\sigma}       
         
\newcommand {\vf }  {\varphi}      
         
      \newcommand {\Om}  {\Omega}
       
\newcommand {\pl}   {\partial}     
\newcommand   {\ex}{{\sf\,e}}            
\newcommand {\one}  {1\!\!1}
\newcommand   {\const}{{\sf\,const}}     
            \renewcommand   {\P}{{\sf P}}
   
\newcommand {\ME}  {{\mathbb E}}   
   \newcommand {\MH}  {{\mathbb H}}
\newcommand {\MM}  {{\mathbb M}}   
   \newcommand {\MP}  {{\mathbb P}}
\newcommand {\MR}  {{\mathbb R}}   
\newcommand {\MS}  {{\mathbb S}}
\newcommand {\MU}  {{\mathbb U}}

   \newcommand {\Sb}  {{\textsc{b}}}

\newcommand {\CC }  {{\cal C}}
   
\newcommand {\Gu}  {\mathfrak{u}}   
   
\newcommand {\Sw}  {{\textsc{w}}}   
   \newcommand {\Sz}  {{\textsc{z}}}

\begin{document}
\title     {On geometric interpretation of the Berry phase}
\author    {M. O. Katanaev
            \thanks{E-mail: katanaev@mi.ras.ru}\\ \\
            \sl Steklov Mathematical Institute,\\
            \sl ul.~Gubkina, 8, Moscow, 119991, Russia}
\date      {\today}
\maketitle
\begin{abstract}
A geometric interpretation of the Berry phase and its Wilczek--Zee non-Abelian
generalization are given in terms of connections on principal fiber bundles. It
is demonstrated that a principal fiber bundle can be trivial in all cases, while
the connection and its holonomy group are nontrivial. Therefore, the main role
is played by geometric rather than topological effects.
\end{abstract}
\section{Introduction}
Berry's phase \cite{Berry84} attracts much interest of theoreticians and
experimenters for a long time. The interest is due to two circumstances. First,
the nontrivial geometric object -- the $\MU(1)$-connection -- arises naturally
when solving the Schr\"odinger equation in nonrelativistic quantum mechanics.
Second, there is a widespread opinion in gauge field theory that only gauge
field strength rather than gauge potentials themselves which are not gauge
invariant can produce the observable effects. Contrary to this opinion,
M.~Berry demonstrated that the integral of a gauge field along a closed loop
could produce observable effects. This conclusion was soon confirmed
experimentally.

The notion of the Berry phase was generalized to the non-Abelian case
corresponding to degenerate energy levels of a Hamiltonian by Wilczek and Zee
\cite{WilZee84}. In this case, non-Abelian $\MU(r)$-gauge fields naturally arise
when solving the Schr\"odinger equation.

In all cases mentioned above, the observable effects are produced by elements of
the holonomy group of corresponding connections. There is no disagreement at
this point. However, there is no common opinion on the geometric interpretation.
B.~Simon \cite{Simon83} and other authors considered the gauge field as a
connection on an associated fiber bundle. Since in general the typical fiber of
the associated fiber bundle is an infinite dimensional Hilbert space, specific
difficulties arose. Important and interesting constructions connected to
characteristic classes are related to the existence of global sections of
associated fiber bundles rather than to the Berry phase itself. Moreover,
definite topological obstructions arose for the existence of global sections.
Therefore, the judgement that the Berry phase has its origin in topology is
widespread in the literature.

In the present paper, a geometric interpretation of the Berry phase in terms of
the connection theory on a principal fiber bundle is given. There are no
difficulties related to infinite dimensional manifolds in this approach, because
typical fibers are $\MU(1)$ or $\MS\MU(r)$ groups which are finite-dimensional
Lie groups. It is demonstrated that the principal fiber bundle can be trivial
while the connection arising on it has generally a nontrivial holonomy group and
therefore leads to observable effects. As a consequence, the Berry phase has its
origin in geometry rather then in topology. Moreover, the existence of global
sections on associated fiber bundles is not a necessary condition. If a global
section is absent, then the local connection forms are defined on a coordinate
covering of the base of the principal fiber bundle defining the unique
connection on the principal fiber bundle up to an isomorphism.
\section{Abelian case: nondegenerate state                       \label{sberph}}
We describe the problem considered by M.~Berry \cite{Berry84} in its simplest
case.

Let the Hilbert space $\MH$ of a quantum mechanical system be finite dimensional
and the Hamiltonian $H=H(\lm)$ depend sufficiently smoothly on a point of a
manifold $\lm\in\MM$ of dimension $\dim\MM=n$. If we choose a coordinate
neighborhood $\MU\subset\MM$ on $\MM$, then the Hamiltonian will depend on $n$
parameters $\lm^k$, $k=1,\dotsc,n$, (coordinates of a point $\lm$). Assume that
the position of point $\lm$ on $\MM$ depends on time $t$ according to a given
prescription, i.e., the Hamiltonian depends on a curve $\lm(t)$,
$t\in[0,\infty]$. Parameterization of the curve by a semiinfinite interval
corresponds to the adiabatic limit \cite{BorFoc28,Messia62} (see also
\cite{Katana11BR}) where $t\to\infty$. We assume also that the Hamiltonian
depends on time only through the point $\lm(t)\in\MM$.

We consider the eigenvalue problem
\begin{equation*}
  H\phi=E\phi,~~~~E=\const,
\end{equation*}
where $\phi\in\MH$ for all $\lm\in\MM$. Suppose there exists nondegenerate
energy eigenvalue $E$ which depends sufficiently smooth on $\lm\in\MM$. The
eigenfunction $\phi(\lm)$ is also assumed to be a sufficiently smooth function
on $\lm$. Without loss of generality, we suppose that the eigenfunction $\phi$
is normalized to unity, $(\phi,\phi)=1$. Then it is unique up to multiplication
on a phase factor which can be $\lm$-dependent. We fix somehow this phase
factor.

Then we solve the Cauchy problem for the Schr\"odinger equation
\cite{Schrod26A,Schrod26B}
\begin{equation}                                                  \label{escheq}
  i\hbar\frac{\pl\psi}{\pl t}=H\psi,
\end{equation}
where $\hbar$ is the Plank constant, with the initial condition
\begin{equation}                                                  \label{einval}
  \psi|_{t=0}=\phi_0,
\end{equation}
where $\phi_0:=\phi\big(\lm(0)\big)$. We set $\hbar=1$ and denote the partial
derivative with respect to time by dot, $\dot\psi:=\pl_t\psi$.

In the adiabatic approximation, a quantum system during evolution remains in
the eigenstate corresponding to energy level $E(\lm)$. Therefore, we seek
solution in the form
\begin{equation*}
  \psi=\ex^{i\Theta}\phi,
\end{equation*}
where $\Theta=\Theta(t)$ is some unknown function on time. Then the
Schr\"odinger equation implies equation for the phase
\begin{equation*}
  \dot\Theta=i(\phi,\dot\phi)-E
\end{equation*}
with the initial condition
$\Theta|_{t=0}=0$. Since $\dot\phi=\dot\lm^k\pl_k\phi$, the solution to the
Cauchy problem for Eq.(\ref{escheq}) is
\begin{equation}                                                  \label{esothe}
  \Theta=\int_0^t\!\! dt\dot\lm^k A_k-\int_0^t\!\! ds E\big(\lm(s)\big)
  =\int_{\lm(0)}^{\lm(t)}\!\! d\lm^k A_k-\int_0^t\!\! ds E\big(\lm(s)\big),
\end{equation}
where we introduced notation
\begin{equation}                                                  \label{edeloc}
  A_k(\lm):=i(\phi,\pl_k\phi)
\end{equation}
and the integral over $\lm$ is taken along the curve $\lm(t)$.

Thus, integral (\ref{esothe}) in the adiabatic approximation yields the solution
of the Cauchy problem for Schr\"odinger equation (\ref{escheq}) with initial
condition (\ref{einval}). The first term in Eq.(\ref{esothe}) is called the
geometric or Berry phase, and the second term is called the dynamical phase.

Note that components (\ref{edeloc}) are real because of normalization of the
wave function. Indeed, differentiation of the normalization condition
$(\phi,\phi)=1$ yields the equality
\begin{equation*}
  (\pl_k\phi,\phi)+(\phi,\pl_k\phi)=(\phi,\pl_k\phi)^\dagger+(\phi,\pl_k\phi)=0.
\end{equation*}
It implies that components (\ref{edeloc}) and hence the Berry phase are real.

We consider now a set of closed curves $\lm\in\Om(\MM,\lm_0)$ on a parameter
manifold $\MM$ with the beginning and end at the point $\lm_0\in\MM$. Then the
total change of the phase of the wave function is equal to the integral
\begin{equation*}
  \Theta=\Theta_\Sb-\int_0^\infty\!\!\!dtE\big(\lm(t)\big),
\end{equation*}
where
\begin{equation}                                                  \label{ebphas}
  \Theta_\Sb=\oint_{\g}\! d\lm^k A_k.
\end{equation}
The dynamical part of the wave function phase diverges. However, we observe in
experiments a difference in phases of two eigenvectors with the same dynamic
phase which is determined by the Berry phase. Therefore, we consider the Berry
phase in more detail.

The Berry phase (\ref{ebphas}) has simple geometric interpretation, namely, we
have a principal fiber bundle $\MP\big(\MM,\pi,\MU(1)\big)$ whose
base is the parameter manifold $\lm\in\MM$, the structure group is $\MU(1)$
(phase of the state vector $\ex^{i\Theta}$), and $\pi:~\MP\rightarrow\MM$ is the
projection \cite{KobNom6369}. The vector in the Hilbert space $\phi\in\MH$
represents a local cross section of the associated fiber bundle
$\ME\big(\MM,\pi_\ME,\MH,\MU(1),\MP\big)$ whose typical fiber is the Hilbert
space $\MH$ and $\pi_\ME:~\ME\rightarrow\MM$ is the projection.

Consider a change in the local cross section of the associated bundle which is
produced by multiplication of a vector in the Hilbert space on a phase factor
(vertical automorphism)
\begin{equation*}
  \phi'=\ex^{ia}\phi,
\end{equation*}
where $a=a(\lm)\in\CC^2(\MM)$ is an arbitrary doubly differentiable function.
Then components (\ref{edeloc}) are transformed according to the rule
\begin{equation*}
  A'_k=A_k-\pl_ka.
\end{equation*}
Comparing this rule with the transformation of components of a local connection
form \cite{KobNom6369}, wee see that the fields $A_k(\lm)$ can be interpreted as
components of a local connection form for the $\MU(1)$ group. In other words,
$A_k(\lm)$ is a gauge field for the one-dimensional unitary group $\MU(1)$. If
the base of the associated fiber bundle
$\ME\big(\MM,\pi_\ME,\MH,\MU(1),\MP\big)$ is covered by some set of coordinate
charts, $\MM=\cup_j\MU_j$, then a set of sections given on each coordinate chart
$\MU_j$ defines a family of local connection forms on the principal fiber bundle
$\MP\big(\MM,\pi,\MU(1)\big)$. A family of local connection forms $d\lm^kA_k$
defines the unique connection on $\MP$ up to an isomorphism \cite{KobNom6369}.

Let us recall the expression for an element of the holonomy group in terms of
the $\P$-exponent \cite{Isham99}. In the present case, the group $\MU(1)$ is
Abelian, and the $\P$-exponent coincides with the conventional exponent.
Therefore, the Berry phase (\ref{ebphas}) defines the element
$\ex^{i\Theta_\Sb}$ of the holonomy group $\Phi(\lm_0,e)\subset\MU(1)$ of the
principal fiber bundle at the point $(\lm_0,e)\in\MP$ corresponding to zero
cross section $\MM\ni\lm\mapsto(\lm,e)\in\MP$ where $\lm_0:=\lm(0)$ and $e$ is
the unit of the structure group $\MU(1)$. The cross section is the zero cross
section because at the initial moment of time the Berry phase vanishes,
$\Theta_\Sb|_{t=0}=0$. The local connection form $d\lm^kA_k$ corresponds also to
zero cross section.

If the base $\MM$ is simply connected, then expression (\ref{ebphas}) for the
Berry phase can be rewritten as a surface integral of the local curvature form.
Using the Stokes formula, we obtain the following expression:
\begin{equation}                                                  \label{ebersu}
  \Theta_\Sb=\frac12\iint_S d\lm^k\wedge d\lm^l F_{kl},
\end{equation}
where $S$ is a surface in $\MM$ with the boundary $\g\in\Om(\MM,\lm_0)$ and
$F_{kl}=\pl_k A_l-\pl_l A_k$ are components of the local curvature form
(gauge field strength). If the base $\MM$ is not simply connected, then the
expression for the Berry phase as a surface integral (\ref{ebersu}) is valid
only for those curves which are contractible to a point.
\subsection{Spin $1/2$ particle in the magnetic field}
As an example, we calculate the Berry phase for a spin 1/2 particle in an
external homogeneous magnetic field. This example is a particular case of a
particle of arbitrary spin in an external homogeneous magnetic field
\cite{Berry84}. In nonrelativistic quantum mechanics, a spin 1/2 particle is
described by a two-component wave function
\begin{equation*}
  \psi=\begin{pmatrix} \psi_+ \\ \psi_- \end{pmatrix}.
\end{equation*}
We assume that it is located in the Euclidean space $\MR^3$ with a given
homogeneous magnetic field. Let the strength of the magnetic field $H^k(t)$,
$k=1,2,3$, do not depend on space point but change in time $t$ according to some
prescribed fashion. For simplicity, we disregard also the kinetic energy of a
particle and assume that other fields are absent. In this case, the Hilbert
space $\MH$ is two-dimensional, and the Hamiltonian of a particle consists of
one term which is equal to the interaction term of magnetic momentum of a
particle with external magnetic field (for example, see \cite{Fock76,Messia62}),
\begin{equation*}
  H=-\mu H^k\s_k,
\end{equation*}
where $\s_k$ are the Pauli matrices and $\mu$ is the magneton (dimensional
constant which is equal to the ratio of the magnetic momentum of the particle
to its spin). To write the Hamiltonian in the form considered above, we
introduce new variables $\lm^k=-\mu H^k$. Then the Hamiltonian is
\begin{equation}                                                  \label{ehaspo}
  H=\lm^k\s_k=\begin{pmatrix}\lm^3 & ~\lm^- \\ \lm^+ & -\lm^3 \end{pmatrix},
\end{equation}
where $\lm^\pm=\lm^1\pm i\lm^2$.

Eigenvalues of Hamiltonian (\ref{ehaspo}) are found from the equation
\begin{equation*}
  \det(H-E\one)=0,
\end{equation*}
which has two real roots
\begin{equation}                                                  \label{ehaigm}
  E_\pm=\pm|\lm|,
\end{equation}
where
\begin{equation*}
  |\lm|=\sqrt{(\lm^1)^2+(\lm^2)^2+(\lm^3)^2}
\end{equation*}
is the length of the vector $\lm=\lbrace\lm^k\rbrace\in\MR^3$. It can be easily
shown, that the equation for eigenfunctions
\begin{equation*}
  H\phi_\pm=E_\pm\phi_\pm,
\end{equation*}
has two solutions
\begin{equation}                                                  \label{eigfuh}
  \phi_\pm=\frac1{\sqrt{2|\lm|}}
  \begin{pmatrix}\pm\displaystyle\frac{\lm^-}{\sqrt{|\lm|\mp\lm^3}} \\
  \\~~{\sqrt{|\lm|\mp\lm^3}} \end{pmatrix}.
\end{equation}
The factor in the expression derived is chosen in such a way that the
eigenfunctions are normalized on unity
\begin{equation*}
  (\phi_\pm,\phi_\pm)=1.
\end{equation*}
Thus Hamiltonian (\ref{ehaspo}) for the 1/2 spin particle in the external
homogeneous magnetic field has two nondegenerate eigenstates (\ref{eigfuh})
corresponding to energy levels (\ref{ehaigm}).

For further calculations in the parameter space $\lm\in\MR^3$, it is convenient
to introduce spherical coordinates $|\lm|,\theta,\vf$:
\begin{align*}
  \lm^1&=|\lm|\sin\theta\cos\vf,
\\
  \lm^2&=|\lm|\sin\theta\sin\vf,
\\
  \lm^3&=|\lm|\cos\theta.
\end{align*}
Then the eigenfunctions assume the form
\begin{equation*}
  \phi_+=\begin{pmatrix} \cos\frac\theta2\ex^{-i\vf} \\ \\
  \!\!\!\sin\frac\theta2 \end{pmatrix},~~~~
  \phi_-=\begin{pmatrix} -\sin\frac\theta2\ex^{-i\vf} \\ \\
  \!\!\!\cos\frac\theta2 \end{pmatrix}.
\end{equation*}

Admit that the experimenter observing the particle varies differentiably
the homogeneous magnetic field with time. That is, the parameters $\lm^k(t)$ in
the Hamiltonian depend differentiably time. Assume also that the particle was in
the state $\phi_+$ at the initial moment of time $t=0$. The corresponding
solution of Schr\"odinger equation (\ref{escheq}) in the adiabatic
approximation is
\begin{equation*}
  \psi=\ex^{i\Theta}\phi_+,
\end{equation*}
where the phase $\Theta$ is defined by Eq.(\ref{esothe}). Components of the
local connection form $A_k=i(\phi_+,\pl_k\phi_+)$ for the eigenstate $\phi_+$
are easily calculated
\begin{equation}                                                  \label{ecomam}
  A_{|\lm|}=0,~~~~A_\theta=0,~~~~A_\vf=\cos^2\frac\theta2.
\end{equation}
The respective local curvature form has only two nonzero components:
\begin{equation*}
  F_{\theta\vf}=-F_{\vf\theta}=-\frac12\sin\theta.
\end{equation*}
Now we calculate the Berry phase for a closed curve in the parameter space
$\g=\lm(t)\in\MM$:
\begin{equation}                                                  \label{eberyp}
\begin{split}
  \Theta_\Sb&=\oint_\g d\lm^k A_k
  =\frac12\iint_S d\lm^k\wedge d\lm^l F_{kl}=
\\
  &=\iint_S d\theta\wedge d\vf F_{\theta\vf}
  =-\frac12\iint_S d\theta\wedge d\vf\sin\theta=-\frac12\Om(\g),
\end{split}
\end{equation}
where $S$ is a surface in $\MR^3$ with the boundary $\g$ and $\Om(\g)$ is
the solid angle under which the surface $S$ is seen from the origin of the
coordinate system.

If the particle is in the state $\phi_-$ at the initial moment of time, the
calculations are similar. In this case,
\begin{equation*}
  A_{|\lm|}=0,~~~~A_\theta=0,~~~~A_\vf=\sin^2\frac\theta2,
\end{equation*}
and components of the local curvature form differ by the sign:
\begin{equation*}
  F_{\theta\vf}=-F_{\vf\theta}=\frac12\sin\theta.
\end{equation*}
Therefore, the Berry phase differs also only by the sign.

Thus, if the particle was in one of the states $\phi_\pm$ at the initial moment
of time, then after variation of the homogeneous magnetic field along closed
curve $\lm(t)$, its wave function acquires the phase factor whose geometrical
part is
\begin{equation}                                                  \label{eberco}
  \Theta_{\Sb\pm}=\mp\frac12\Om(\g),
\end{equation}
where $\Om(\g)$ is the solid angle under which the closed contour $\g$ is seen
from the origin of coordinates. This result is valid in the adiabatic
approximation when parameters $\lm(t)$ change slowly with time. Expression
(\ref{eberco}) for the Berry phase was confirmed experimentally \cite{BitDub87}
for the scattering of polarized neutrons in a spiral magnetic field.

The homogeneous magnetic field in the above-considered example can have an
arbitrary direction and magnitude. Therefore, the base $\MM$ of the principal
fiber bundle $\MP\big(\MM,\pi,\MU(1)\big)$ coincides with the Euclidean space
$\MM=\MR^3$. Hence, the principal fiber bundle $\MP$ is trivial,
$\MP\approx\MR^3\times\MU(1)$. For the Berry phase, the connection on this fiber
bundle is given by the section of the associated fiber bundle, for example,
$\phi_+$ which is obtained by solving the Schr\"odinger equation. It is easily
checked that this section (\ref{eigfuh}) has a singularity on the positive
half-axis $\lm^3\ge0$. The components of local connection form (\ref{ecomam})
in the Cartesian coordinates have the form
\begin{equation}                                                  \label{elocco}
\begin{split}
  A_1&=\frac{\pl\vf}{\pl\lm^1}A_\vf
  =-\frac{\sin\vf\cos\frac\theta2}{2|\lm|\sin\frac\theta2},
\\
  A_2&=\frac{\pl\vf}{\pl\lm^2}A_\vf
  =~~\frac{\cos\vf\cos\frac\theta2}{2|\lm|\sin\frac\theta2},
\\
  A_3&=\frac{\pl\vf}{\pl\lm_3}A_\vf=0.
\end{split}
\end{equation}
At this point, we are obliged to use Cartesian system of coordinates, because
the spherical system of coordinates are singular on the $\lm^3$ axis and is
unsuitable for an analysis of singularities located here. We see that the
components of the local connection form are singular on the positive half-axis
$\lm^3\ge0$ together with the vector $\phi_+$. Now we calculate the components
of the local form of the curvature tensor. All its components are nonzero:
\begin{align*}
  F_{12}&=-F_{21}=-\frac{\cos\theta}{2|\lm|^2},
\\
  F_{13}&=-F_{31}=~~\frac{\sin\theta\sin\vf}{2|\lm|^2},
\\
  F_{23}&=-F_{32}=-\frac{\sin\theta\cos\vf}{2|\lm|^2}.
\end{align*}
Finally, we calculate the square of the curvature tensor which is the
geometric invariant:
\begin{equation*}
  F^2=2\big(F_{12}^2+F_{13}^2+F_{23}^2\big)=\frac1{2|\lm|^4}.
\end{equation*}
Thus, the curvature form is singular only at the origin of coordinates.

Let us return to our principal fiber bundle $\MR^3\times\MU(1)$. The local
connection form (\ref{elocco}) is not defined on it, because it is singular on
the half-axis $\lm^3\ge0$ which we denote $\lbrace\lm^3_+\rbrace$. Hence, to
construct a principle fiber bundle with the given connection, we have to remove
the inverse image $\pi^{-1}\big(\lbrace \lm^3_+\rbrace)$ where
$\pi:~\MR^3\times\MU(1)\rightarrow\MR^3$ is the natural projection. As a result,
we get the trivial fiber bundle
$\big(\MR^3\setminus\lbrace \lm^3_+\rbrace\big)\times\MU(1)$ which is the
subbundle on the initial one. Local connection form (\ref{elocco}) is smooth
on this principal fiber bundle.

We may get another way out. Since the magnetic field is external, then we can
assume that it varies, for eaxample, in the half space $\MR^3_+$ defined by the
inequality $\lm_1>0$. The corresponding principal fiber bundle is trivial
$\MP\approx\MR^3_+\times\MU(1)$, because the half space $\MR^3_+$ is
diffeomorphic to all Euclidean space $\MR^3$. In this case, no problem
arise with the definition of the connection, because local connection form
(\ref{elocco}) is smooth. At the same time, previous expression (\ref{eberco})
for the Berry phase is valid.

Thus, the Berry phase is the geometric rather then topological notion, because
the topology of the principal fiber bundle is trivial. It arises due to
nontrivial connection defined by cross sections of the associated fiber bundle.
\subsection{Non-Abelian case: degenerate state                   \label{swizee}}
The notion of the Berry phase was generalized to the case when energy levels of
the Hamiltonian are degenerate \cite{WilZee84}. In this case, the principle
fiber bundle $\MP\big(\MM,\pi,\MU(r)\big)$ with the structure group $\MU(r)$,
where $r$ is the number of independent eigenfunctions corresponding to the
degenerate energy level $E$, appears when solving the Schr\"odinger equation.
Here we describe this construction in detail.

We suppose that the Hamiltonian of a quantum system depends on a point of some
manifold $\lm(t)\in\MM$ as was assumed earlier. Let $E$ be a degenerate
eigenvalue of a Hamiltonian $H$ with $r$ independent eigenfunctions $\phi^a$,
$a=1,\dotsc,r$,
\begin{equation*}
  H\phi^a=E\phi^a
\end{equation*}
for all moments of time. We assume that $E(\lm)$ and $\phi^a(\lm)$ are
differentiable functions at a point $\lm$ of the manifold, and the number of
eigenfunctions $r$ does not change in time.

The eigenfunctions can be chosen orthonormalized
\begin{equation*}                                                 \label{eortnf}
  (\phi^a,\phi_b)=\dl^a_b,
\end{equation*}
where $\dl^b_a$ is the Kronecker symbol and $\phi_b=\phi^a\dl_{ab}$. We search
for solution $\psi^a$ of the Cauchy problem for Schr\"odinger equation
(\ref{escheq}) with the initial condition
\begin{equation*}
  \psi^a(0)=\psi^a_0=\phi^a\big(\lm(0)\big).
  \end{equation*}
That is, the system is in one of the eigenstates $\phi^a$ at the initial moment
of time. In the adiabatic approximation, solution $\psi^a$ is the eigenstate
of the Hamiltonian $H(\lm)$ corresponding to the energy value $E(\lm)$ for all
moments of time. Therefore it can be decomposed with respect to eigenfunctions
of the degenerate state
\begin{equation}                                                  \label{esutra}
  \psi^a=U^{-1}{}_b^a\phi^b,
\end{equation}
where  $U(\lm)\in\MU(r)$ is some unitary matrix which depends differentiably on
the point $\lm\in\MM$.

The unitarity of the matrix $U$ is dictated by the following circumstance.
Consider solutions $\psi^a$ for all values of index $a=1,\dotsc,r$.
Differentiating the scalar product $(\psi^a,\psi_b)$ with respect to time and
using the Schr\"odinger equation, we obtain equation
\begin{equation*}
  \frac\pl{\pl t}(\psi^a,\psi_b)=-i(\psi^aH,\psi_b)+i(\psi^a,H\psi_b)=0.
\end{equation*}
The last equality follows from the self-adjointness of the Hamiltonian. As a
consequence, if vectors $\psi_0^a=\phi^a\big(\lm(0)\big)$ are orthonormalized
at the initial moment of time, then the corresponding solutions of the
Schr\"odinger equation remain orthonormalized for all subsequent moments of
time. Hence the matrix $U$ in decomposition(\ref{esutra}) is unitary.

The Schr\"odinger equation for solution (\ref{esutra}) is reduced to equation
\begin{equation*}
  i\dot U^{-1}{}_c^b\phi^c+iU^{-1}{}_c^b\dot\phi^c
  =HU^{-1}{}_c^b\phi^c.
\end{equation*}
Let us take the scalar product of the left and right hand sides with $\phi_a$.
As a result, we derive equation for the unitary matrix
\begin{equation}                                                  \label{euneqm}
  \dot U^{-1}{}_a^b=\dot\lm^k U^{-1}{}_c^bA_{ka}{}^c-iEU^{-1}{}_a^b,
\end{equation}
where we have introduced the following notation:
\begin{equation}                                                  \label{edegau}
  A_{ka}{}^c:=-(\pl_k\phi^c,\phi_a).
\end{equation}

Orthonormality of eigenfunctions $\phi^a$ implies antiunitarity of components
$A_{ka}{}^b$ for all $k=1,\dotsc,n$ if indices $a$ and $b$ are considered as
matrix ones. Indeed, differentiating the orthonormality condition
$(\phi^b,\phi_a)=\dl_a^b$ we obtain equality
\begin{equation*}
  (\pl_k\phi^b,\phi_a)+(\phi^b,\pl_k\phi_a)=(\phi^a,\pl_k\phi_b)^\dagger
  +(\phi^b,\pl_k\phi_a)=0.
\end{equation*}
That is, matrices $A_k$ are antiunitary and therefore belong to the Lie algebra
$\Gu(r)$. Consequently, the matrices $A_k$ define 1-forms in some neighborhood
$\MU\subset\MM$ with values in the Lie algebra, as components of a local
connection form.

The initial condition for the unitary matrix has the form
\begin{equation*}
  U^{-1}{}_a^b|_{t=0}=\dl_a^b.
\end{equation*}
The solution of the Cauchy problem for Eq.(\ref{euneqm}) can be written as the
$\P$-product
\begin{equation}                                                  \label{ebenab}
\begin{split}
  U^{-1}(t)&=\P\exp\left(\int_0^t\!\! ds\dot\lm^k(s)A_k(s)
  -i\int_0^t\!\! dsE\big(\lm(s)\big)\right)=
\\
  &=\P\exp \left(\int_{\lm(0)}^{\lm(t)}\!\!d\lm^k A_k\right)
  \times\exp\left(-i\int_0^t\!\! dsE\big(\lm(s)\big)\right),
\end{split}
\end{equation}
where we have omitted matrix indices for simplicity.

The first factor is the generalization of the Berry phase to the case of
degenerate states, and the second factor is the dynamical phase. The dynamical
phase has the same form as for the nondegenerate state.

The first factor in solution (\ref{ebenab}) represent of Wilczek--Zee unitary
matrix
\begin{equation}                                                  \label{ewizem}
  U^{-1}_{\Sw\Sz}=\P\exp \left(\int_{\lm(0)}^{\lm(t)}\!\!\!d\lm^k A_k\right),
\end{equation}
which can be given the following geometric interpretation. We have the principal
fiber bundle $\MP\big(\MM,\pi,\MU(r)\big)$ with the structure group $\MU(r)$
(transformation (\ref{esutra})). The set of eigenfunctions $\phi^a$ is the cross
section of the associated fiber bundle
$\ME\big(\MM,\pi_\ME,\MH^r,\MU(r),\MP\big)$ with the typical fiber being
the tensor product of Hilbert spaces
\begin{equation*}
  \MH^r:=\underbrace{\MH\otimes\dotsc\otimes\MH}_r.
\end{equation*}
Under the vertical automorphism given by the unitary matrix $U(\lm)\in\MU(r)$,
\begin{equation*}
  \phi^{\prime a}=U^{-1}{}_b^a\phi^b,~~~~~\phi'_a=U_a^b\phi_b,
\end{equation*}
fields (\ref{edegau}) transform according to the rule
\begin{equation}                                                  \label{elotrs}
  A'_k=U^{-1}A_kU+U^{-1}\pl_k U,
\end{equation}
where we have omitted matrix indices. It implies that the fields $A_k$ can be
interpreted as components of the local connection form or Yang--Mills fields.
A set of these components given on a coordinate covering of the base $\MM$
defines uniquely the connection on the principal fiber bundle
$\MP\big(\MM,\pi,\MU(r)\big)$.

If the path is closed, $\lm\in\Om(\MM,\lm_0)$, then the unitary Wilczek--Zee
matrix (\ref{ewizem}) represents the element of the holonomy group
$U^{-1}_{\Sw\Sz}\in\Phi(\lm_0,e)$ at the point $(\lm_0,e)\in\MP$ corresponding
to the zero cross section $\MM\ni\lm\mapsto(\lm,e)\in\MP$ where $\lm_0:=\lm(0)$
and $e$ is the unity of the structure group $\MU(r)$.

So the principal fiber bundle $\MP\big(\MM,\pi,\MU(r)\big)$ arises in the
case of a degenerate energy level of the Hamiltonian. In the above-considered
case, the base $\MM$ is the parameter manifold $\lm\in\MM$ with the Hamiltonian
being dependent on its point. We suppose that this manifold is finite
dimensional. The structure group is the unitary group $\MU(r)$ which is also
finite dimensional. The connection on the principal fiber bundle is defined by
the cross sections of the associated bundle
$\ME\big(\MM,\pi_\ME,\MH^r,\MU(r),\MP\big)$. Generally, the typical fiber of the
associated fiber bundle can be infinite dimensional Hilbert space $\MH^r$. In
the present paper we do not consider infinite dimensional manifolds to avoid
difficulties which can arise \cite{Zharin08}. Nevertheless, in our case
everything that is needed is the transformation formulas for components of
local connection form (\ref{elotrs}) which can be easily checked in each
particular case. If the associated bundle is not diffeomorphic to the direct
product $\MM\times\MH^r$, then the state of a quantum system is given by a
family of local cross sections on a coordinate covering of the base $\MM$. It
defines a family of local connection forms (\ref{edegau}). In its turn, the
family of local connection forms defines a connection on the principal fiber
bundle $\MP\big(\MM,\pi,\MU(r)\big)$ uniquely up to an isomorphism.

We see once again that principal and associated fiber bundles can be trivial or
not depending on the problem under consideration. The connection on the
principal fiber bundle $\MP\big(\MM,\pi,\MU(r)\big)$ can be nontrivial and imply
nontrivial Wilczek--Zee matrix (\ref{ewizem}) describing parallel transport
of fibers along a path on the base $\lm(t)\in\MM$ even for trivial bundles. This
observation confirms its geometric rather than topological origin. For closed
paths $\lm\in\Om(\MM,\lm_0)$ with the beginning and end at a point
$\lm_0\in\MM$, the Wilczek--Zee matrix defines the element of the holonomy group
$U_{\Sw\Sz}\in\Phi(\lm_0,e)\subset\MU(r)$.

For simplicity, we assumed the Hilbert space for the Berry phase and
Wilczek--Zee matrix to be finite dimensional. This assumption can be essentially
relaxed. The formulas obtained are valid for all levels for which the adiabatic
theorem holds, that is, it must be an isolated level with the energy level
separated from all the rest of the spectrum.
\section{Conclusions}
In this paper, we have considered the Berry phase and its non-Abelian
generalization by Wilczek and Zee. These effects are demonstrated to be the
consequences of nontrivial connections on the principal fiber bundles which
define the nontrivial holonomy group. At the same time, the topology of the
principal fiber bundle can be trivial. Therefore, the considered effects are not
topological as they are often called in modern physical literature but rather
geometric effects.

The interpretation proposed in the paper contains nothing except differential
geometric notions. In a geometric interpretation of mathematical physics models,
one has to take into account that a connection exists on any principal fiber
bundle independently of the topology of the base \cite{KobNom6369}. Moreover, if
a family of local connection forms is given on an arbitrary closed submanifold
of the base of some principal fiber bundle, then the corresponding connection
can always be extended to the whole principal fiber bundle. This can be done in
many ways. A connection defines the holonomy group which is nontrivial in the
general case.

In experiments on testing the existence of the Berry phase, the observable
effects are produced not by the whole holonomy group but a fixed element of the
holonomy group which depends on the connection and the closed contour. The
topology of the base can be trivial or not, it does not play any role. If the
topology is trivial, then the contour can be contracted to a point. The effect
disappears in this case, because the corresponding element of the holonomy group
tends to the unity element, and this is quite natural from the physical point of
view.

A connection on the principal fiber bundle defines connections on all fiber
bundles which are associated with it. In particular, if the typical fiber is an
infinitely dimensional Hilbert space, then the connection is also defined. At
present, the interpretation of the Berry phase, as a rule, is reduced to
consideration of a connection on an associated fiber bundle, and this forces one
to consider infinite dimensional manifolds and to take into account the related
subtleties. From our point of view, the interpretation of the geometric effects
in terms of connections on principal fiber bundles is simpler and more natural.

The author is grateful to I.~V.~Volovich and D.~V.~Treschev for the discussions
of the paper and useful comments. The work was supported in part by the Russian
Foundation of Basic Research (Grants No.\ 11-01-00828-a and 11-01-12114-ofi\_m),
the Program for Supporting Leading Scientific Schools
(Grant No.\ NSh-7675.2010.1), and the program ``Modern problems of theoretical
mathematics'' by the Russian Academy of Sciences.


\begin{thebibliography}{10}

\bibitem{Berry84}
M.~V. Berry.
\newblock Quantal phase factors accompanying adiabatic changes.
\newblock {\em Proc.\ Roy.\ Soc.\ London}, A392(1802):45--57, 1984.

\bibitem{WilZee84}
F.~Wilczek and A.~Zee.
\newblock Appearance of gauge structure in simple dynamical systems.
\newblock {\em Phys.\ Rev.\ Lett.}, 52:2111, 1984.

\bibitem{Simon83}
B.~Simon.
\newblock Holonomy, the quantum adiabatic theorem, and berry's phase.
\newblock {\em Phys.\ Rev.\ Lett.}, 51(24):2167--2170, 1983.

\bibitem{BorFoc28}
M.~Born and V.~Fock.
\newblock{Beweis des Adiabatensatzes},
\newblock{Z.\ Phys.}, 51:165--180, 1928.
\newblock{English translation in {\it ``V.A. Fock -- Selected Works: Quantum
Mechanics and Quantum Field Theory''} ed.\ by L.D.~Faddeev, L.A.~Khalfin,
I.V.~Komarov. Chapman \& Hall/CRC, Boca Raton, 2004.}

\bibitem{Messia62}
A.~Messiah.
\newblock {\em Quantum Mechanics}, volume~2.
\newblock North Holland, Amsterdam, 1962.

\bibitem{Katana11BR}
M.~O.~Katanaev.
\newblock Adiabatic theorem for finite dimensional quantum mechanical systems.
\newblock {\em Izv. Vuzov. Phisics}.

\bibitem{Schrod26A}
E.~Schr\"odinger.
\newblock {Quantizierung als Eigenwertproblem (Erste Mitteilung)}.
\newblock {\em Ann.\ Phys.\ Leipzig}, 79(4):361--376, 1926.

\bibitem{Schrod26B}
E.~Schr\"odinger.
\newblock {Quantizierung als Eigenwertproblem (Zweite Mitteilung)}.
\newblock {\em Ann.\ Phys.\ Leipzig}, 79(6):489--527, 1926.

\bibitem{KobNom6369}
S.~Kobayashi and K.~Nomizu.
\newblock {\em Foundations of differential geometry}, volume 1, 2.
\newblock Interscience publishers, New York -- London, 1963.

\bibitem{Isham99}
C.~J. Isham.
\newblock {\em Modern Differential Geometry}.
\newblock World Scientific, Singapore, 1999.

\bibitem{Fock76}
V.~A.~Fock.
\newblock {\em Foundations of quantum mechanics.}
\newblock Nauka, Moscow, 2nd ed., 1976 [in Russian].

\bibitem{BitDub87}
T.~Bitter and D.~Dubbers.
\newblock Manifestation of Berry's topological phase in neutron spin rotation.
\newblock {\em Phys.\ Rev.\ Lett.}, 59:251--254, 1987.

\bibitem{Zharin08}
V.~V.~Zharinov.
\newblock {\em Algebro-geometric foundations of mathematical physics}.
\newblock Steclov Mathmatical Institute, Moscow, 2008.

\end{thebibliography}
\end{document}